\documentclass[reprint,
 amsmath,amssymb,
 aps,
]{revtex4-1}

\usepackage{graphicx}% Include figure files
\usepackage{dcolumn}% Align table columns on decimal point
\usepackage{bm}% bold math
\begin{document}

\preprint{APS/123-QED}

\title{Expansion Dynamics of Two Dimensional Extended Bose-Hubbard Model }

\author{Sevda AKTA\c{S}}
 \email{sevdaaktas@selcuk.edu.tr}
\author{\"{U}lfet ATAV}%
 \email{uatav@selcuk.edu.tr}

\affiliation{%
 Physics Department, Sel\c{c}uk University, 42075 Konya, TURKEY
}%

\date{\today}% It is always \today, today,

\begin{abstract}
We study the expansion dynamics of harmonically trapped bosons in a two-dimensional lattice within the extended Bose-Hubbard model. We evaluate the dynamics of the system following a sudden removal of the confining potential, starting with a cloud mostly in $n=1$ Mott state.  We show that the nearest neighbour interactions have a strong influence on the dynamics of ultracold bosons on an optical lattice. Also we conclude that validity of the widely used contact potential approximation is questionable in the presence of Feschbach resonances. 

\begin{description}
\item[PACS numbers]
 \pacs{} 03.75.Kk, 67.85.−d, 05.30.Jp
\end{description}
\end{abstract}

\maketitle

%\tableofcontents

Recent developments in experimental techniques have allowed precise control over the properties of ultracold atomic gases in traps and optical lattices, turning them into a simulation tool for quantum many body systems and allowing experimental investigation of static and dynamic behaviour of clean and low dimensional systems. Thanks to the Feshbach resonance that it is possible to control the strength and sign of the interactions between the particles of a many body system which had been a long lasting dream for many people. Interactions between the particles of a many body system determine both the static and dynamic behaviour of the system. Various properties of the interactions such as range, strength, sign etc. are important in determining the behaviour of the system. Static properties and the phase structure of ultracold atomic gases on optical lattices have been thoroughly investigated \cite{trefzger08,menotti07,iskin12,hen10,danshita07,beibing11,iskin13,xianlong06}. Experimental studies on dynamical properties of ultracold atomic gases have also revealed many peculiar behaviour which deserves special attention and further theoretical investigation \cite{pezze04,ott04,strohmaier07,schneider12,ronzheimer13}. Extensive theoretical studies on dynamics of ultracold atoms in 1D have yielded results consistent with experiments \cite{li13,vidmar13}.
However some experimental results in 2D  still lacks sufficient explanation deserves further theoretical investigation \cite{schneider12,ronzheimer13}.

The effect of the interactions on expansion dynamics of initially confined fermionic \cite{schneider12} and bosonic \cite{ronzheimer13} gases on a two dimensional (2D) optical lattice was studied by changing the interactions via a Feshbach resonance. It was observed that qualitative behaviour of the expansion dynamics is independent of the sign of the interactions for the fermions suddenly released from an isotropic harmonic trap \cite{schneider12}. For small interaction strengths (small s-wave scattering lengths) the atomic clouds released from the isotropic trap changes its symmetry from circular (trap) to square (optical lattice) as it expands. However, for larger interaction strengths core of the atomic cloud does not expand at all and preserves its circular symmetry. This bimodal behaviour was  also observed in the free expansion of bosonic atoms released from an initial Mott state and again the behaviour was the same for both attractive or repulsive interactions \cite{ronzheimer13}.   

Motivated by the experimental results mentioned above, we study the effect of the interparticle interactions on the expansion properties of bosons on a 2D optical lattice.  Jreissaty et al. have studied expansion of bosons released from a harmonic trap on an optical lattice by using a Bose-Hubbard hamiltonian with on-site interactions only \cite{jreissaty11, jreissaty13} where they have calculated the time evolution of the cloud in both the real and momentum space.  Their model have predicted a seperation of the initial cloud into slowly and rapidly expanding two clouds. However, the strong localization observed in 2D experimental studies \cite{ronzheimer13} were not predicted by this model \cite{jreissaty13}. Therefore, it appears that the standard Bose Hubbard model can not capture the essence of physics behind the experimental observation of zero expansion velocities in the presence of interactions \cite{ronzheimer13}.

Considering only the on-site interactions is a common assumption in the theory of ultracold atomic gases in optical lattices. However, Duan L.M. have shown that for a broad Feshbach resonance nearest neighbour coupling rates can be significantly large compared to atom tunnelling rates and thus should not be neglected \cite{duan05}. Also, using a Rydberg dressing can lead to effectively long ranged interactions between ultracold atoms \cite{li12}. Therefore, a more realistic description of the interacting ultracold atoms near a Feshbach resonance should include at least the nearest neighbour interactions.

We consider a system of bosons released from a harmonic trap in a 2D optical lattice, assuming the interactions between the atoms have a significant long range part. Such a system can be described by extended Bose-Hubbard Hamiltonian
\begin{eqnarray}
\hat{H}=-J \sum_{<i,j>} \left ( \hat{a}_i^\dag \hat{a}_j+ \hat{a}_j^\dag \hat{a}_i \right )+\frac{U_0}{2}\sum_{i} \hat{n}_i(\hat{n}_i-1) \nonumber \\ 
 + \frac{U_1}{2}\sum_{<i,j>} \hat{n}_i\hat{n}_j + \sum_{i}(V_i-\mu) \hat{n}_i
\label{Hamil}
\end{eqnarray}
where $ \hat{a}_i$ ($ \hat{a}_i^\dag$) is the boson creation (annihilation) operator at a given lattice site $i$, and $\hat{n}_i$ is the particle number operator. The first two terms in the Hamiltonian are the usual kinetic energy and on site interaction terms that define the well known Hubbard model in the homogeneous case with $J$ being the hopping amplitude between neighbouring sites and $U_0$  being the on-site interaction strength. The third term describes the nearest neighbour interactions and the last term comes from the trapping and chemical potentials. The summation index $<i,j>$ means that the summation is to be performed over nearest neighbour sites.
\begin{figure}[b]
\includegraphics[width=3 in]{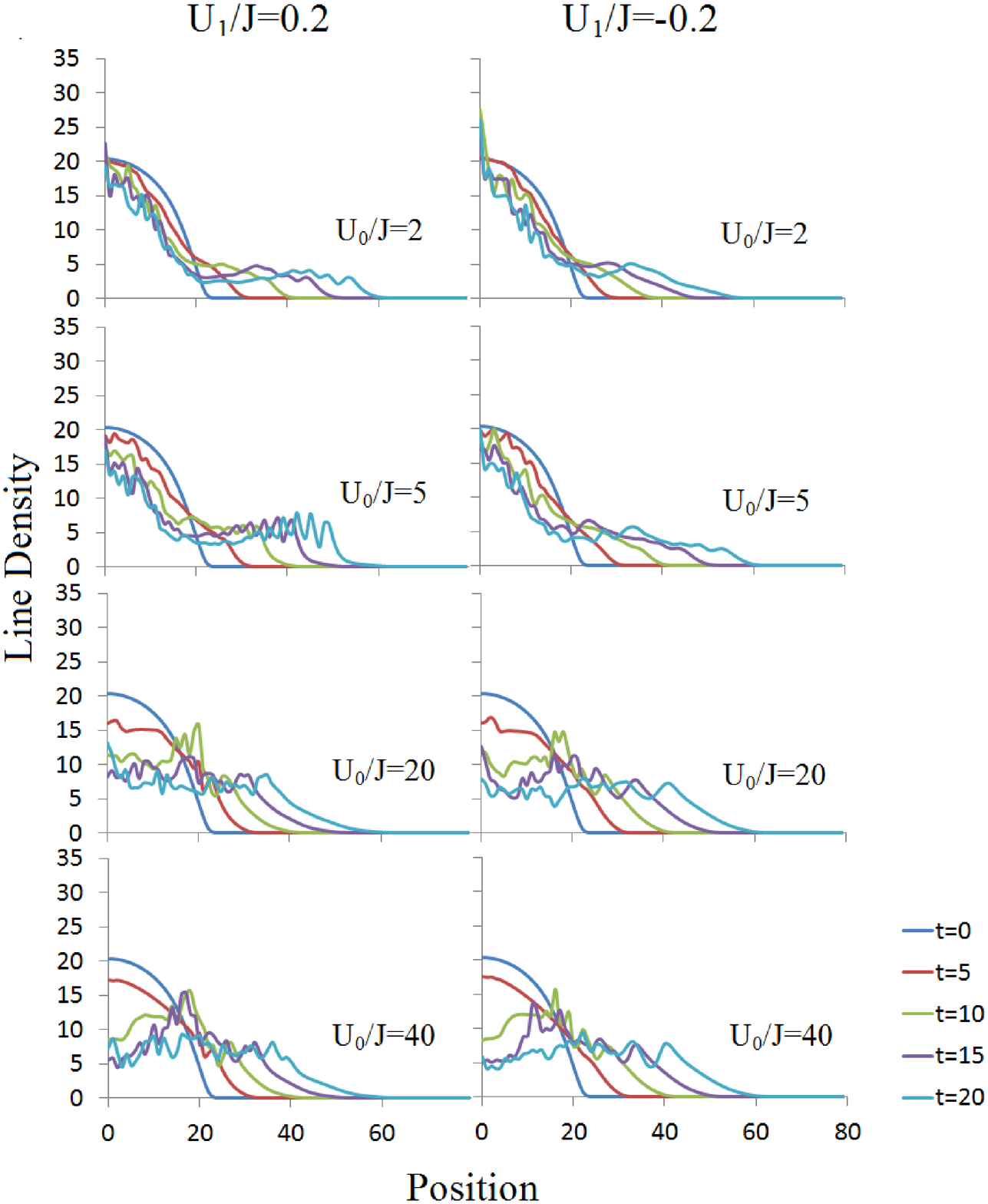}
% Here is how to import EPS art
\caption{ \label{fig:DP1}Line densities of the expanding cloud for $U_1/J=0.2, -0.2 $  at various times after  turning off the trap.}
\end{figure}
\begin{figure}[b]
\includegraphics[width=3 in]{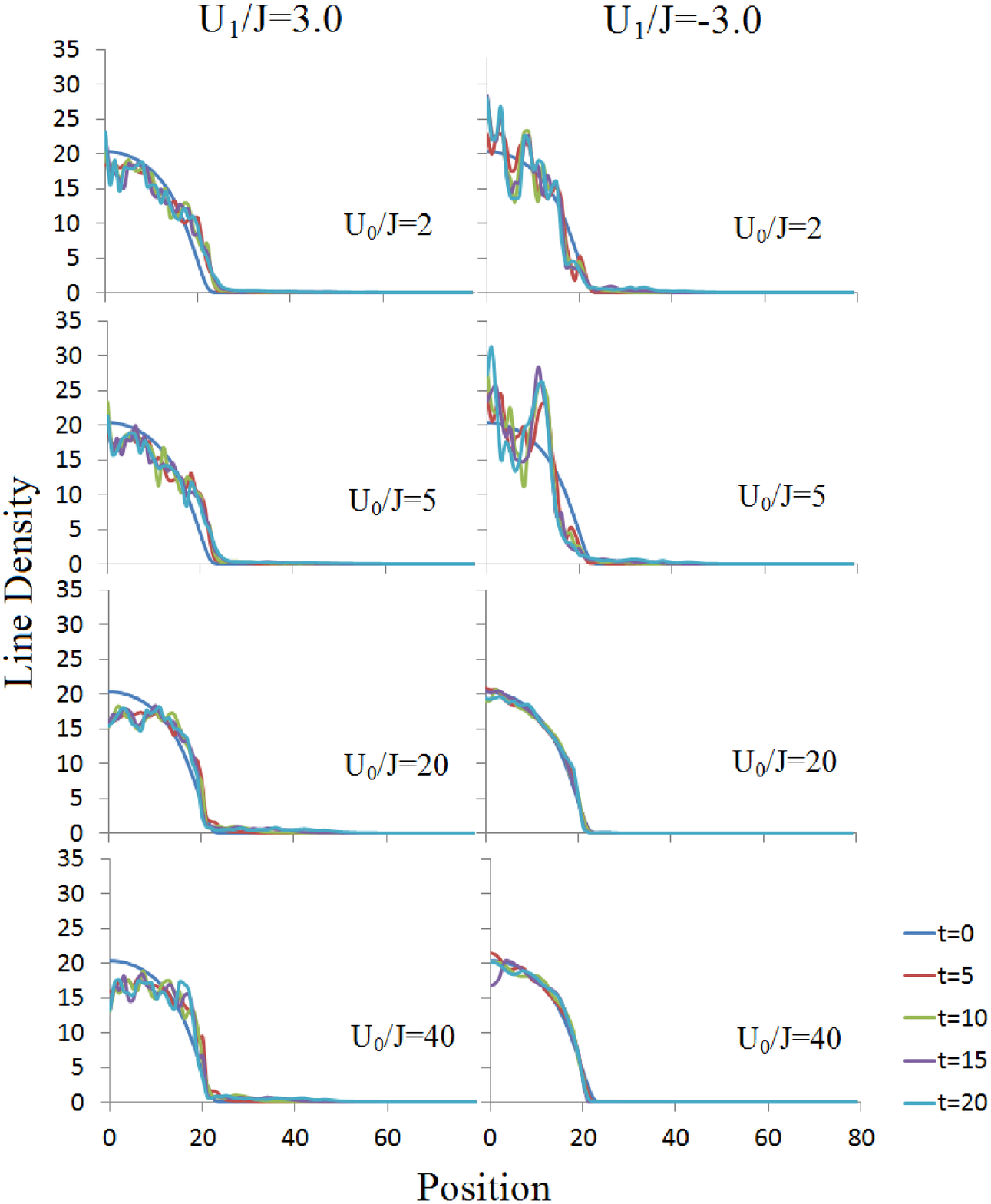}
% Here is how to import EPS art
\caption{ \label{fig:DP2} Line densities of the expanding cloud for $U_1/J=3.0, -3.0$, at various times after  turning off the trap.}
\end{figure}
We use the Gutzwiller ansatz  $ \lvert \psi \rangle = \prod_i \sum_{n=0}^{n_{max}} {f_n^i \lvert n \rangle_i} $ for the many body wave function
and the equations of motion for the probability amplitudes $f_n^i$ are  
\begin{eqnarray}
i \hslash \frac {d f_n^i}{dt}=-J \left[ \bar{\phi}_i \sqrt{n}f_{n-1}^i +\bar{\phi^*}_i \sqrt{n+1}f_{n+1}^i\right] \nonumber \\
+n \left[ \frac {U_0}{2} (n-1) + \sum_{<j,i>} {\langle \hat n_j \rangle} +V_i-\mu \right] f_n^i  
\label{time}
\end{eqnarray}
\noindent where $ \phi_j=\langle\hat{a}_j \rangle $ is the order parameter for site $j$ and $\bar\phi_i=\sum_{<j,i>} \phi_j $.
To be consistent with the experimental work of Ronzheimer et al.\cite{ronzheimer13} we start from an initial state which is mostly in the $(n=1)$ Mott insulator state. Such an initial state was obtained by the approach described by Jreissaty et. al \cite{jreissaty11}. Then, the cloud is allowed to expand by removing the trapping potential with a simultaneous quench of the  interaction potential. After getting the initial ground state we turn off the trap and follow the time evolution of the system by integrating Eq.(\ref{time}) using a fourth-order Runge-Kutta method. 

In Fig. \ref{fig:DP1} line densities of the cloud for weak nearest neighbour interaction strengths ($U_1/J=0.2 , -0.2$) are given. For small on site interaction strengths the cloud clearly seperates into two parts: the central part slowly melts down as the surrounding cloud rapidly expands. As the on site interaction strength increases, the central part more rapidly melts down, and the density of the expanding part increases however the edge velocity of the expanding cloud seems to be insensitive to the changes in $U_0$. Our results for $U_1=0$ case is qualitatively similar to the ones presented in Fig. \ref{fig:DP1}. Jreissaty et al.\cite{jreissaty11} had previously considered $U_1=0$ case obtaining results consistent with the present study in the absence of nearest neighbour interactions. Also, it is interesting to note that expansion dynamics are very similar for attractive and repulsive cases. This sign reversal symmetry was experimentally observed and theoretically discussed by Schneider et al. \cite{schneider12}. 
\begin{figure*}[bt]
\includegraphics[width=6 in]{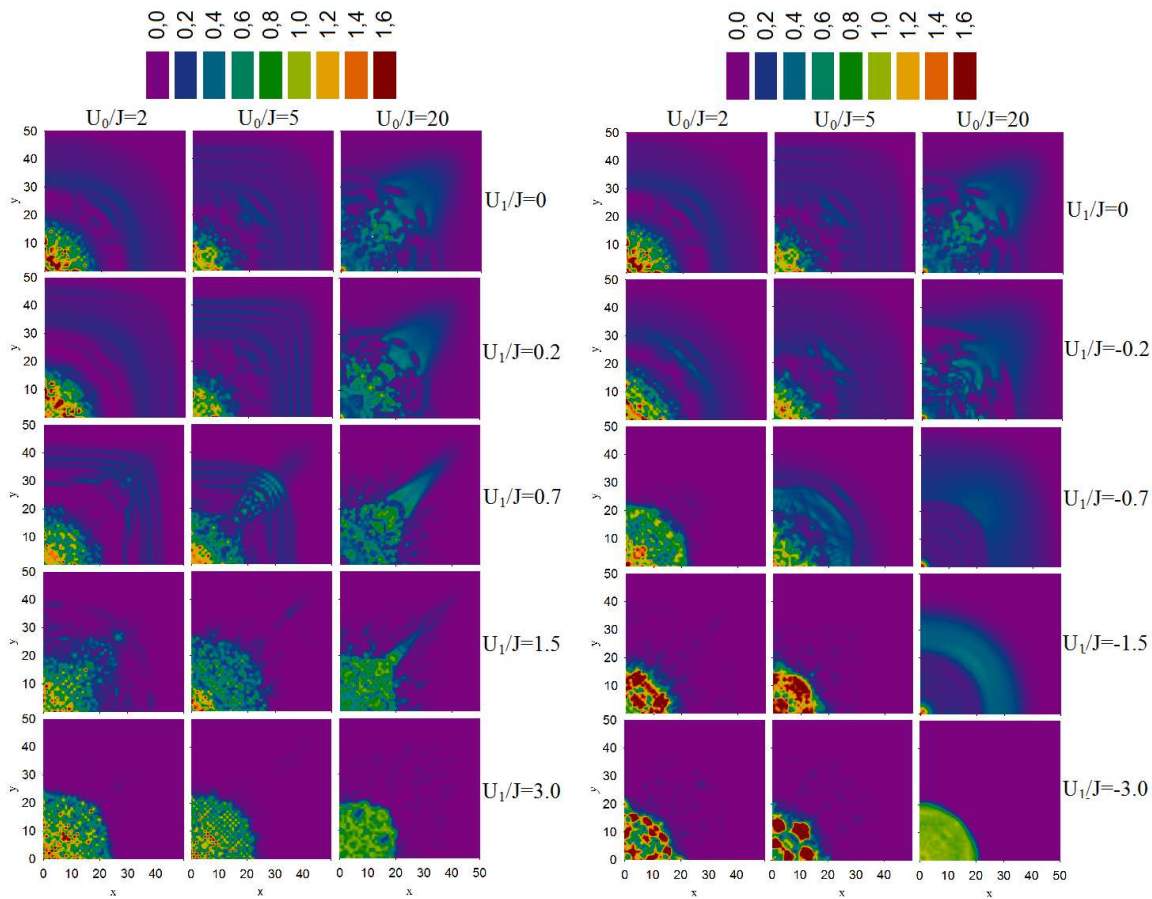}
% Here is how to import EPS art
\caption{ \label{fig:v2} Density distributions of the expanding clouds at time $t = 15 \hslash /J$ for various interaction parameters.  Due to the symmetry of the system only a quarter of the expanding cloud is presented.  }
\end{figure*}
Fig. \ref{fig:DP2} presents the line densities of the released cloud for nearest neighbour interaction strengths  ($U_1/J=3.0, -3.0$). The edge of the density profiles are almost fixed and the cloud does not expand. This localization behaviour of the cloud is independent of the on site interaction strength.      
On the other hand, the structure of the cloud strongly depends on the value of $U_0/J$. For small $U_0/J$ values there are strong fluctuations in the density profile while for large $U_0/J$ values these fluctuations are suppressed due to the enhanced thermalization. Comparing Figs. \ref{fig:DP1} and \ref{fig:DP2} one may conclude that the expansion dynamics is dominantly controlled by the nearest neighbour interactions. Even though the self trapping of the cloud occurs for both repulsive and attractive interactions, the shape and texture of the cloud shows  a clear distinction between the two cases. Especially for high $U_0$ values the cloud seems to exhibit a circular symmetry for attractive nearest neighbour interactions and a square symmetry for the repulsive case.
   
The texture of expanding cloud also strongly depends on the interaction parameters. Density distribution of the expanding clouds are given in Fig. \ref{fig:v2} at time $t=15 \hslash / J$ after the release of the cloud for various interaction strengths. For weak nearest neighbour interactions repulsive and attractive cases are very similar. However as the magnitude $U_1/J$ approaches to unity the behaviour of attractive and repulsive cases differ. When $|U_1/J>1|$ the cloud is more homogeneous for repulsive interactions while high density lumps are formed for attractive nearest neighbour interactions. However, if $U_0/U_1 \gg 1$ high density lumps disappears as the repulsive on site interactions dominates. 

Expansion velocities can be used to quantify the discusssion of expansion dynamics. Some recent experimental work use half width at half maximum (HWHM) as the core radius and determine core expansion velocity as the time rate of change of HWHM. However there are some drawbacks of using this measure. Strong oscillations seen in Figs. \ref{fig:DP1} and \ref{fig:DP2} in the line density profiles might result in erroneous determination of the core radius. The seperation of the cloud into slowly and rapidly expanding parts poses another difficulty. Therefore we use alternative measures for the expansion velocities.
\begin{figure}[t]
\includegraphics[width=3 in]{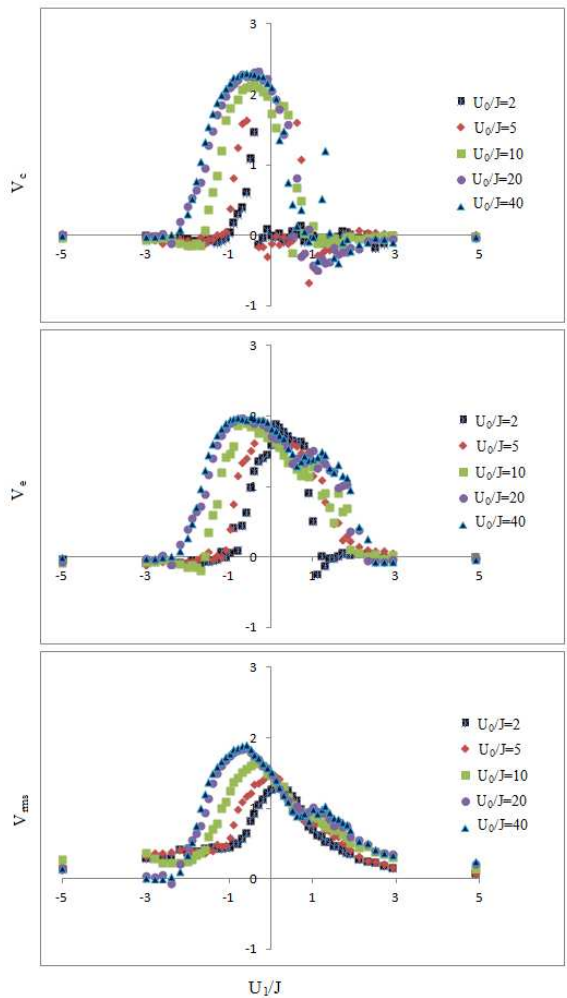}
% Here is how to import EPS art
\caption{ \label{fig:velo} Expansion velocities of the cloud with respect to the nearest neighbour interaction strength for various on site interaction strengths. Top panel shows the core expansion velocity $V_c$, middle panel shows the edge expansion velocity $V_{edge}$ and the bottom panel shows the rms expansion velocity $V_{rms}$. The velocities are given in units of $\frac{d}{\hslash/J}$.}
\end{figure}
To measure the core radius, first we define the correlation coefficient
\begin{equation}
C(R)= \frac {\int {\langle \hat{n}(r)\rangle  \theta(R-r)d\tau}} {\sqrt{\int{ {\langle \hat{n}(r)\rangle}^2 d\tau} \int{(\theta(R-r))^2 d\tau} }}
\label{corr}
\end{equation}								
\noindent where $\theta(x)$ is the Heaviside step function. The R value corresponding to the maximum of $C(R)$ is taken as the core radius, $R_c$. Then core expansion velocity is $V_c =d R_c /dt$. Note that for lattice systems Eq. (\ref{corr}) becomes
\begin{equation}
C(R)= \frac {\sum_i {\langle \hat{n}(r_i)\rangle  \theta(R-r_i)}} {\sqrt{\sum_i    { {\langle \hat{n}(r_i)\rangle}^2 } \sum_i {(\theta(R-r_i))^2 } }}
\label{corr_latt}
\end{equation}								
An alternative measure of the expansion velocity is the time rate of change of root mean square radius ($R_{rms}$) of the cloud which is defined for lattice systems as
\begin{equation}
R_{rms}=\sqrt { \frac{1}{N} \sum_i {r_i^2 \langle \hat{n}(r_i)\rangle  }}
\label{rms}
\end{equation}								
Then $rms$ expansion velocity is $V_rms =d R_rms /dt$.
The core and $rms$ expansion velocities are useful in describing the dynamics of the cloud as a whole, however when the cloud seperates into slowly and rapidly expanding parts one may need other descriptions for the expansion velocity. The edge of the cloud can be used as a measure of the extent of rapidly expanding part. We determine the edge of the cloud as follows. First we calculate the line densities, then as we move from outside of the cloud towards the center the first point at which the line density reaches to 10\% of the maximum line density is recorded as the edge radius $R_{edge}$. Then edge expansion velocity is $V_{edge} =d R_{edge} /dt$.    

Fig. \ref{fig:velo} shows the above defined expansion velocities for various on site and nearest neighbour interaction strengths. The expansion velocities of the cloud shows qualitatively the same behaviour for all on-site interaction strengths, however the maximum expansion velocity shifts to the left as on-site interaction strength increases. The strength of the nearest neighbour interactions, on the other hand, have a dramatic influence on the dynamics of the cloud and all expansion velocities quickly drops to zero for $|U_1/J|>1$. In Fig. \ref{fig:velo} expansion velocities are presented for only $|U_1/J|<5$ because the calculated expansion velocities are essentially zero for the nearest neighbour interaction strengths beyond $|U_1/J|=5$. This result is consistent with the peculiar self trapping behaviour observed in  experimental studies \cite{schneider12,ronzheimer13}. 

For relatively small on site interaction strengths $(U_0/J=2, 5)$ the core does (top panel in Fig. \ref{fig:velo}) not expand even for very small nearest neighbour interactions. Instead the core slowly melts down and the surrounding low density cloud expands with $V_{edge}$.
 
Even though the self trapping behaviour observed for both strongly attractive and strongly repulsive nearest neighbour interactions, our results show that the dynamics of the cloud is not exactly symmetrical for attractive and repulsive interactions. The maximum of the expansion velocity is observed for some negative nearest neighbor interactions, and the position of this maximum shifts to the left ( towards more attractive interactions) with increasing on site interactions. Because the experimental results of Schneider et al. \cite{schneider12} and Ronzheimer et al. \cite{ronzheimer13} lacks the detail when the interaction strengths and the hopping amplitude are of the same order of magnitude we can not compare this asymmetry in detail with the experimental results. However, the experimental density distributions presented in Fig. 3 of Schneider et al. are very similar for the pair $U/J=-1.0$, $U/J=0.5$ and for the pair $U/J=-1.7$, $U/J=1.3$ indicating existence of such an asymmetry.

The strong agreement between the experimental observations and the results of the present study leads us to two important conclusions: First, the dynamics of ultracold bosons on an optical lattice strongly depends on the nearest neighbour interactions. In contrast to the on-site interactions, the effect of the nearest neighbour interactions is beyond naive expectations and a small change in the nearest neighbour interaction may cause dramatic changes in the dynamical behaviour.  
Second, the consistence of our results with experimental studies of ultracold atoms \cite{schneider12,ronzheimer13} implies that effective nearest neighbour interactions are considerably large for these experimental setups, and this in turn implies that the interactions enhanced by a Feschbach resonance have significant long range parts. Therefore, one should reconsider the validity of widely used contact potential approximation while working on ultracold atoms near a Feschbach resonance. For an acceptable description of the dynamics of ultracold atoms one must take the long range interactions into account even when these interactions are considerably weak.

\end{document}